\documentclass[aps,twocolumn,showpacs,preprintnumbers,nofootinbib,prl,superscriptaddress,groupedaddress,10pt]{revtex4-1}
\usepackage[utf8]{inputenc}
\usepackage{graphicx,amssymb,amsmath,amsthm,amsfonts,epsfig,times,natbib}

\usepackage[usenames,dvipsnames]{color}
\usepackage{epstopdf}
\usepackage{amsmath,amssymb}
\usepackage{tensor}
\usepackage{mathtools}
\usepackage{amsbsy}
\usepackage{bm,url}
\usepackage[linktocpage]{hyperref}

\usepackage[scaled]{beramono}
\usepackage[T1]{fontenc}

\definecolor{oxfordblue}{rgb}{0.0, 0.13, 0.28}
\definecolor{burgundy}{rgb}{0.5, 0.0, 0.13}
\definecolor{darkolivegreen}{rgb}{0.33, 0.42, 0.18}
\definecolor{darkblue}{rgb}{0,0,0.5}
\definecolor{richcarmine}{rgb}{0.84, 0.0, 0.25}
\definecolor{darkblue}{rgb}{0,0,0.5}
\definecolor{venetianred}{rgb}{0.78, 0.03, 0.08}
\definecolor{skobeloff}{rgb}{0.0, 0.48, 0.45}
\hypersetup{colorlinks=true, citecolor=darkblue, linkcolor=darkblue,
urlcolor = darkblue}

\def\apj{\rm{ApJ}}
\def\apjl{\rm{ApJ}}

\def\aap{\rm{A\&A}}

\def\mnras{\rm{MNRAS}}

\newcommand{\ben}{\begin{enumerate}}
\newcommand{\een}{\end{enumerate}}

\def\be{\begin{equation}}
\def\ee{\end{equation}}
\def\bea{\begin{eqnarray}}
\def\eea{\end{eqnarray}}

\newcommand{\beq}{\begin{eqnarray}}
\newcommand{\eeq}{\end{eqnarray}} 
\newcommand{\ba}{\begin{align}}
\newcommand{\ea}{\end{align}}

\begin{document}

\title{
Stochastic and resolvable gravitational waves from ultralight bosons}

\author{
Richard Brito$^{1}$,
Shrobana Ghosh$^{2}$,
Enrico Barausse$^{3}$,
Emanuele Berti$^{2,4}$,
Vitor Cardoso$^{4,5}$,
Irina Dvorkin$^{3,6}$,
Antoine Klein$^{3}$,
Paolo Pani$^{7,4}$
}
\affiliation{${^1}$ Max Planck Institute for Gravitational Physics (Albert Einstein Institute), Am M\"{u}hlenberg 1, Potsdam-Golm, 14476, Germany}
\affiliation{${^2}$ Department of Physics and Astronomy, The University of 
Mississippi, University, MS 38677, USA}
\affiliation{${^3}$ Institut d'Astrophysique de Paris, Sorbonne
  Universit\'es, UPMC Univ Paris 6 
  \& CNRS, UMR 7095, 98 bis bd Arago, 75014 Paris, France}
\affiliation{${^4}$ CENTRA, Departamento de F\'isica, Instituto Superior
T\'ecnico, Universidade de Lisboa, Avenida Rovisco Pais 1,
1049 Lisboa, Portugal}
\affiliation{${^5}$ Perimeter Institute for Theoretical Physics, 31 Caroline Street North
Waterloo, Ontario N2L 2Y5, Canada}
\affiliation{${^6}$ Institut Lagrange de Paris (ILP), Sorbonne Universit\'es, 98 bis bd Arago, 75014 Paris, France}
\affiliation{${^7}$ Dipartimento di Fisica, ``Sapienza'' Universit\`a di Roma \& Sezione INFN Roma1, Piazzale Aldo Moro 5, 00185, Roma, Italy}

\begin{abstract}
  Ultralight scalar fields around spinning black holes can trigger
  superradiant instabilities, forming a long-lived bosonic condensate
  outside the horizon. We use numerical solutions of the perturbed
  field equations and astrophysical models of massive and stellar-mass
  black hole populations to compute, for the first time, the
  stochastic gravitational-wave background from these sources. In optimistic scenarios the
  background is observable by Advanced LIGO and LISA for field masses
  $m_s$ in the range $\sim [2\times 10^{-13}, 10^{-12}]\,{\rm eV}$ and
  $\sim 5\times[ 10^{-19}, 10^{-16}]\,{\rm eV}$, respectively, and
  it can affect the detectability of resolvable sources. Our estimates suggest that an analysis of the stochastic background limits from LIGO O1 might already be
used to marginally exclude axions with mass $\sim 10^{-12.5}{\rm eV}$. Semicoherent searches with Advanced LIGO (LISA) should
  detect $\sim 15~(5)$ to $200~(40)$ resolvable sources for scalar
  field masses $3\times 10^{-13}$ ($10^{-17}$)~eV. LISA measurements
  of massive BH spins could either rule out bosons in the range $\sim [10^{-18}, 2\times 10^{-13}]$~eV, or measure $m_s$ with ten
  percent accuracy in the range $\sim[10^{-17}, 10^{-13}]$~eV.
\end{abstract}

\maketitle

\noindent{{\bf{\em Introduction.}}}
%
The historical LIGO gravitational wave~(GW)
detections~\cite{Abbott:2016blz,Abbott:2016nmj,Abbott:2017vtc} provide
the strongest evidence to date that astrophysical black holes (BHs)
exist and merge~\cite{Berti:2005ys,Cardoso:2016rao,Berti:2016lat}.
Besides probing the nature of compact objects and testing general
relativity~\cite{Gair:2012nm,Yunes:2013dva,Berti:2015itd,TheLIGOScientific:2016src},
LIGO~\cite{TheLIGOScientific:2014jea} and the space-based detector LISA~\cite{Audley:2017drz} may
revolutionize our understanding of particle physics and dark matter.
Ultralight bosons,
which could be a significant component of dark
matter~\cite{Arvanitaki:2009fg,Essig:2013lka,Marsh:2015xka,Hui:2016ltb},
interact very weakly (if at all) with baryonic matter, but the
equivalence principle implies that their gravitational interaction
should be universal. Low-energy bosons near spinning BHs can trigger a
superradiant instability whenever the boson frequency $\omega_R$
satisfies the superradiant condition $0<\omega_R<m \Omega_{\rm H}$,
where $\Omega_{\rm H}$ is the horizon angular velocity and $m$ is an
azimuthal quantum number, with possible astrophysical
implications~\cite{Press:1972zz,Detweiler:1980uk,Cardoso:2004nk,Brito:2015oca}.

Despite extensive work on massive
spin-0~\cite{Detweiler:1980uk,Zouros:1979iw,Cardoso:2005vk,Dolan:2007mj},
spin-1~\cite{Pani:2012vp,Pani:2012bp,Witek:2012tr,Endlich:2016jgc,East:2017ovw,Baryakhtar:2017ngi}
and spin-2 fields~\cite{Brito:2013wya}, the evolution and the
end-state of the instability are not fully
understood~\cite{Okawa:2014nda,Zilhao:2015tya,Sanchis-Gual:2015lje,Bosch:2016vcp}.
Recent numerical simulations~\cite{East:2017ovw} support the
conclusions of perturbative
studies~\cite{Arvanitaki:2010sy,Brito:2014wla,Arvanitaki:2014wva,Yoshino:2014wwa,Brito:2015oca,Arvanitaki:2016qwi,Baryakhtar:2017ngi}: the BH spins down,
transferring energy and angular momentum to a mostly dipolar boson
condensate until $\omega_R\sim m\Omega_H$. The energy scale is set by
the boson mass $m_s\equiv \mu \hbar$, which implies that
$\omega_R\sim \mu$ and that the instability saturates at
$\mu\sim m\Omega_{\rm H}$ (in units $G=c=1$). The condensate is then
dissipated through the emission of mostly quadrupolar GWs, with
frequency set by $\mu$.  The mechanism is most effective when
the boson's Compton wavelength is comparable to the BH's gravitational radius: 
detailed calculations show that the maximum instability
rate for scalar fields corresponds to
$M\mu\simeq 0.42$~\cite{Dolan:2007mj}. Therefore, the
instability window corresponds to masses
$m_s\sim 10^{-14}$--$10^{-10}$~eV and
$m_s\sim 10^{-19}$--$10^{-15}$~eV for LIGO and LISA BH-boson condensate sources,
respectively~\cite{Brito:2015oca}.
In this work and in a companion paper~\cite{Brito:2017zvb} we argue that GW
detectors can discover new particles beyond the Standard Model or
impose constraints on their masses.

\noindent{{\bf{\em GWs from scalar condensates around BHs.}}}
%
The instability occurs in two stages~\cite{Brito:2014wla}.
In the first (linear) phase the condensate grows on a timescale
$\tau_{\rm inst}\sim M^{-8}\mu^{-9}$ until the superradiant condition is
nearly saturated.  In the second (nonlinear) phase
GW emission governs the evolution of the condensate, which is
dissipated over a timescale $\tau_{\rm GW}$ that depends on its mass
$M_S$ and on the GW emission rate.
These two timescales can be computed analytically when
$M\mu\ll 1$~\cite{Brito:2017zvb}. For small dimensionless BH spins 
$\chi\equiv J/M^2\ll 1$, they read
\begin{eqnarray}
\tau_{\rm inst}& \sim& 0.07 \, \chi^{-1}\left(\frac{M}{10\,M_\odot}\right)  \left(\frac{0.1}{M\mu}\right)^{9} \,{\rm yr} \,, \label{tauSR}\\
\tau_{\rm GW}&\sim& 6\times 10^{4}\, \chi^{-1}\left(\frac{M}{10\,M_\odot}\right)  \left(\frac{0.1}{M\mu}\right)^{15} \,{\rm yr}\,. \label{tauGW}
\end{eqnarray}
These relations (valid for any BH mass) are a good approximation even
when $M\mu$ and $\chi$ are $\sim 1$~\cite{Brito:2017zvb}.
Since $\tau_{\rm GW}\gg\tau_{\rm inst}\gg M$, the condensate has
enough time to grow, and the evolution of the system can be studied in
a quasi-adiabatic approximation~\cite{Brito:2014wla} using Teukolsky's
formalism~\cite{Teukolsky:1973ha,Yoshino:2013ofa}. The field's
stress-energy tensor is typically small,
thus its backreaction is negligible~\cite{Brito:2014wla,East:2017ovw}.


Over the emission timescale (which in most cases is much longer that the observation time $T_{\rm obs}$), the GWs are nearly
monochromatic, with frequency $f_s=\omega_R/\pi\sim \mu/\pi$. As such,
BH-boson condensates are continuous sources, like pulsars for LIGO or verification binaries for LISA.
We conservatively assume that GWs are produced after saturation of the
instability, which leads the BH from an initial state $(M_i,\,J_i)$ to
a final state $(M,\,J)$, and we thus compute the root-mean-square
strain amplitude $h$ using the \emph{final} BH parameters. By
averaging over source and detector orientations we get
\begin{equation}
h=\sqrt{\frac{2}{5\pi}}\frac{GM}{c^2r}\left(\frac{M_S}{M}\right){A(\chi,f_s M)}\,,
\label{strain}
\end{equation}
where $r$ is the (comoving) distance to the source, the masses are in the source frame, and the
dimensionless function $A(\chi,f_s M)$
is computed from BH perturbation
theory~\cite{Yoshino:2013ofa,Brito:2017zvb}. Our results are more accurate
than the analytic approximations
of~\cite{Arvanitaki:2010sy,Brito:2014wla}.  It can be shown that $M_S$
scales linearly with $J_i$~\cite{Brito:2017zvb}, so $h$ also grows with
$J_i$.  For LISA, we also take into account correction factors due to
the detector geometry~\cite{Berti:2004bd}.  
In the detector frame, Eq.~\eqref{strain} still holds if the masses
$M$ and $M_S$ are multiplied by $(1+z)$, $r$ is replaced by
the luminosity distance, and the frequency is replaced by the
detector-frame frequency $f=f_s/(1+z)$.  Nevertheless, one needs to
use detector-frame frequencies when comparing to the detector
sensitivity.

In semicoherent searches of monochromatic sources, the signal is
divided in ${\cal N}$ coherent segments of time length $T_{\rm coh}$,
and we have
$h_{\rm thr}\simeq 25{\cal N}^{-1/4} \sqrt{S_h(f)/T_{\rm coh}}$, where
$h_{\rm thr}$ is the minimum root-mean-square strain amplitude
detectable over the observation time
${\cal N}\times T_{\rm coh}$~\cite{Palomba:2012wn}, and $S_h(f)$ is
the noise power spectral density (PSD) at $f$~\cite{Ruiter:2007xx}.

\begin{figure}[t]
\includegraphics[width=0.48\textwidth]{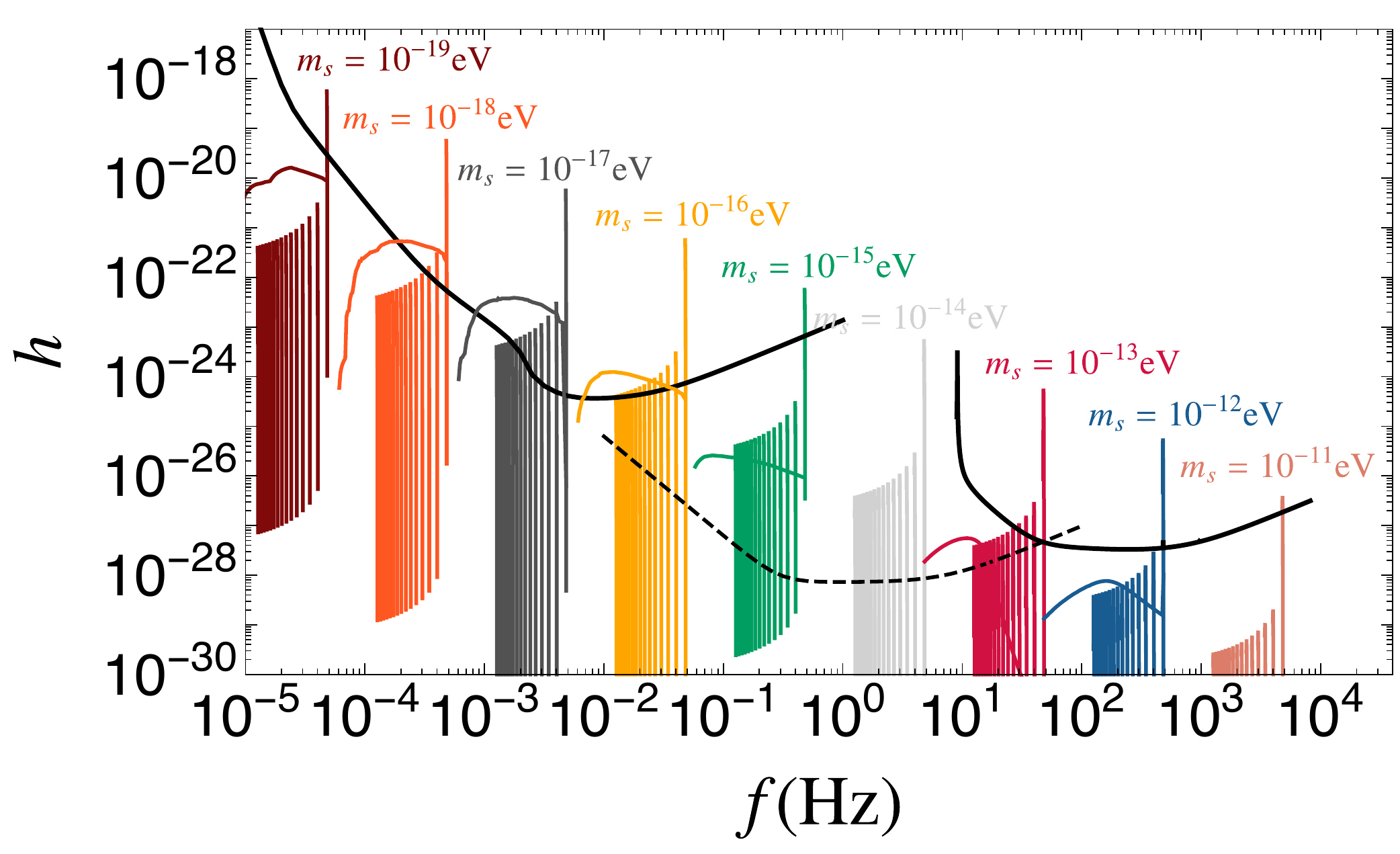}
\caption{GW strain produced by BH-boson condensates compared to the
  Advanced LIGO PSD at design sensitivity~\cite{Aasi:2013wya} and to
  the non-sky averaged LISA PSD~\cite{Audley:2017drz} (black thick
  curves), assuming a coherent observation time of
  $T_{\rm obs}=4\,{\rm yr}$ in both cases. Nearly vertical lines
  represent BHs with initial spin $\chi_i=0.9$. Each line corresponds
  to a single source at redshift $z\in(0.001,3.001)$ (from right to
  left, in steps of $\delta z=0.2$), and different colors correspond
  to different boson masses $m_s$.  Thin lines show the stochastic
  background produced by the whole population of astrophysical BHs
  under optimistic assumptions (cf. main text for details).  The PSD
  of DECIGO~\cite{Kawamura:2006up} (dashed line) is also shown for
  reference.
  \label{sensitivity}}
\end{figure}

In Fig.~\ref{sensitivity} we compare the GW strain of
Eq.~\eqref{strain} with the PSDs of LISA and Advanced LIGO at design
sensitivity.  The GW strain increases almost vertically as
a function of $\omega_R\simeq \mu$ in the superradiant range
$(0,\Omega_H)$.  
Thin solid curves correspond to the stochastic background from
the whole BH population, for a boson mass $m_s$.  This
background produces itself a ``confusion noise'' when
$m_s\approx [10^{-18}, 10^{-16}]$~eV, complicating the detection of
individual sources.
Figure~\ref{sensitivity} suggests that bosons with masses
$10^{-19}\,{\rm eV} \lesssim m_s \lesssim 10^{-11}\,{\rm eV}$ (with a
small gap around $m_s\sim 10^{-14}$~eV, which might be filled by
DECIGO~\cite{Kawamura:2006up}) could be detectable by LIGO and LISA.
Below we quantify this expectation.

%

\noindent{{\bf{\em BH population models.}}}
%
Assessing the detectability of these signals requires
astrophysical models for BH populations.
For LISA sources, the main uncertainties concern the mass and spin
distribution of isolated BHs, the model for their high-redshift seeds, 
and their accretion and merger history.  We adopt the same
populations of~\cite{Klein:2015hvg,Babak:2017tow}, which were based
on the semianalytic galaxy formation calculations
of~\cite{Barausse:2012fy} (see
also~\cite{Sesana:2014bea,newpaper,letter}).  In our {\em optimistic}
model, we use these calculations to infer the redshift-dependent BH
number density ${d^2 n}/(d \log_{10} M d \chi)$.  The spin
distribution is skewed toward $\chi_i\sim 1$, at least at low
masses~\cite{Sesana:2014bea}.
We also adopt {\em less optimistic} and {\em pessimistic} models with
mass function given by Eqs.~(5) and (6) of~\cite{Babak:2017tow} for
$z<3$ and $10^4 M_\odot< M<10^7 M_\odot$, whereas for $M>10^7 M_\odot$
we use a mass distribution with normalization $10$ and $100$ times
lower than the optimistic one. In both the {\em less optimistic} and
{\em pessimistic} models we assume a uniform spin
distribution in the range $\chi_i\in [0,1]$.

The LIGO stochastic GW background comes mostly from extra-galactic
stellar-mass BHs, which were ignored in previous
work~\cite{Arvanitaki:2014wva}. Here we model these sources using the
semianalytic galaxy evolution model of~\cite{2016MNRAS.461.3877D}. The
BH formation rate as a function of mass and redshift reads
\begin{equation}
\frac{d \dot{n}_{\rm eg}}{dM}=\int \textrm{d}{\cal M_\star}\psi[t-\tau({\cal {M_\star}})]\phi({\cal M_\star})\delta[{\cal M_\star}-g^{-1}({M})]\,,
\label{Irina}
\end{equation}
where $\tau({\cal M_\star})$ is the lifetime of a star of mass
${\cal M_\star}$, $\phi({\cal M_\star})$ is the stellar initial mass
function, $\psi(t)$ is the cosmic star formation rate (SFR) density
and $\delta$ is the Dirac delta. 
We fit the cosmic SFR as described
in~\cite{2015MNRAS.447.2575V} and calibrate it to observations of
luminous galaxies~\cite{2011ApJ...737...90B,2013ApJ...770...57B}. We
assume a Salpeter initial mass function
$\phi({\cal M_\star})\propto {\cal
  M_\star}^{-2.35}$~\cite{1955ApJ...121..161S} in the range
${\cal M_\star}\in[0.1-100]\,M_{\odot}$, and take stellar lifetimes
from~\cite{2002A&A...382...28S}. We also follow the production of
metals by stars~\cite{1995ApJS..101..181W} and the resulting
enrichment of the interstellar medium, which affects the metallicity
of subsequent stellar generations. The function $g({\cal M_\star})$ relates the initial stellar mass
$\cal M_\star$ and the BH mass $M$, and 
encodes the BH formation process. In general, the mass of the BH
formed from a star with initial mass ${\cal M_\star}$ depends on the
stellar metallicity \citep{2008NewAR..52..419V} and rotational
velocity \citep{2009A&A...497..243D}, as well as interactions with its
companion if the star belongs to a binary system. We assume that all
stellar-mass BHs are produced from isolated massive stars after core
collapse, and calculate the BH mass for a given ${\cal M_\star}$ and
metallicity using the analytic fits for the ``delayed'' model
of~\cite{2012ApJ...749...91F}. Through the metallicity, the function
$M=g({\cal M_\star})$ is implicitly a function of redshift.
Since this model does not predict the initial BH spins, we
assume a uniform distribution and explore different ranges:
$\chi_i\in [0.8,1]$, $[0.5,1]$, $[0,1]$ and $[0,0.5]$.
%

\begin{figure*}[t]
\begin{center}
\epsfig{file=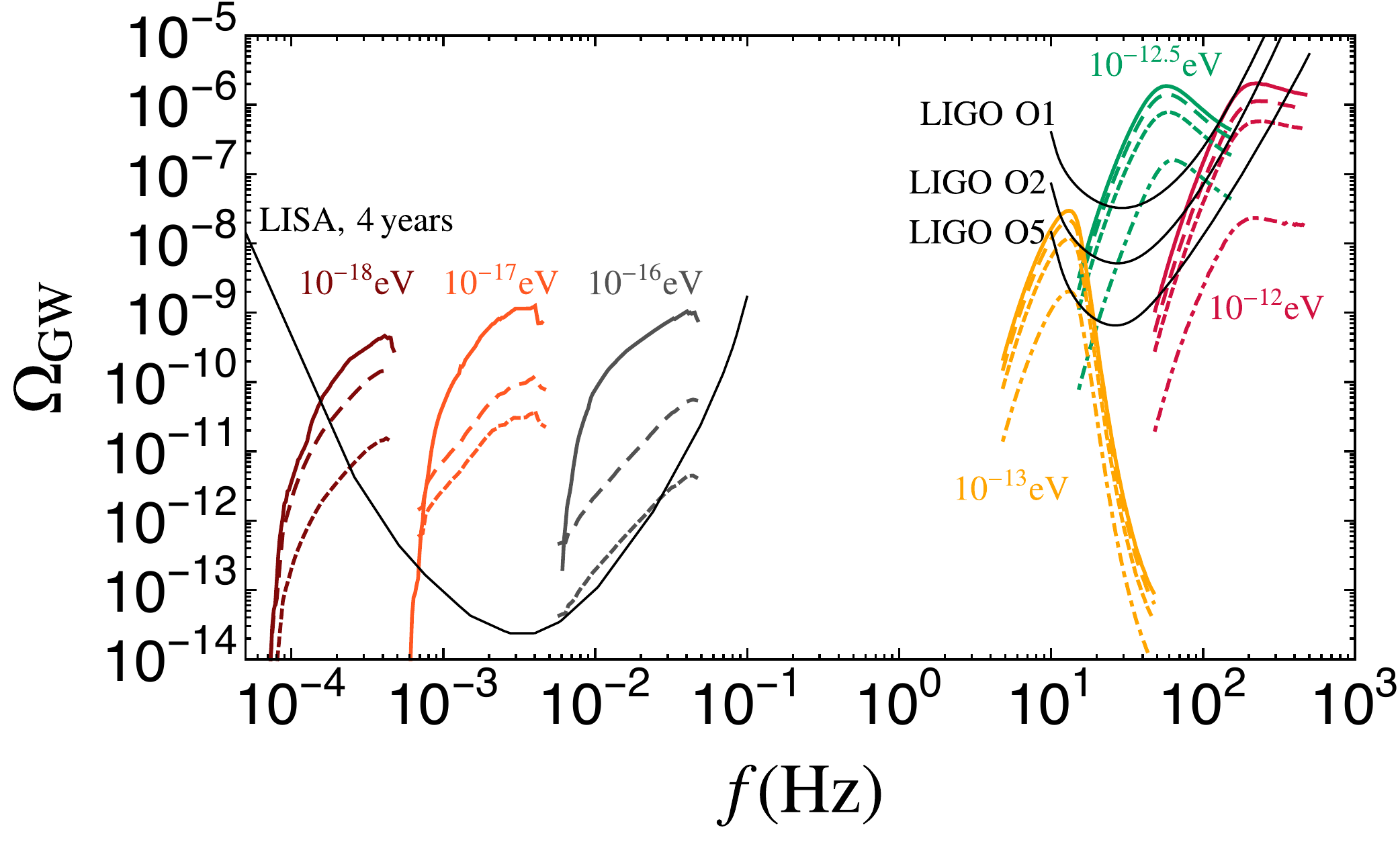,width=0.48\textwidth,angle=0,clip=true}
\epsfig{file=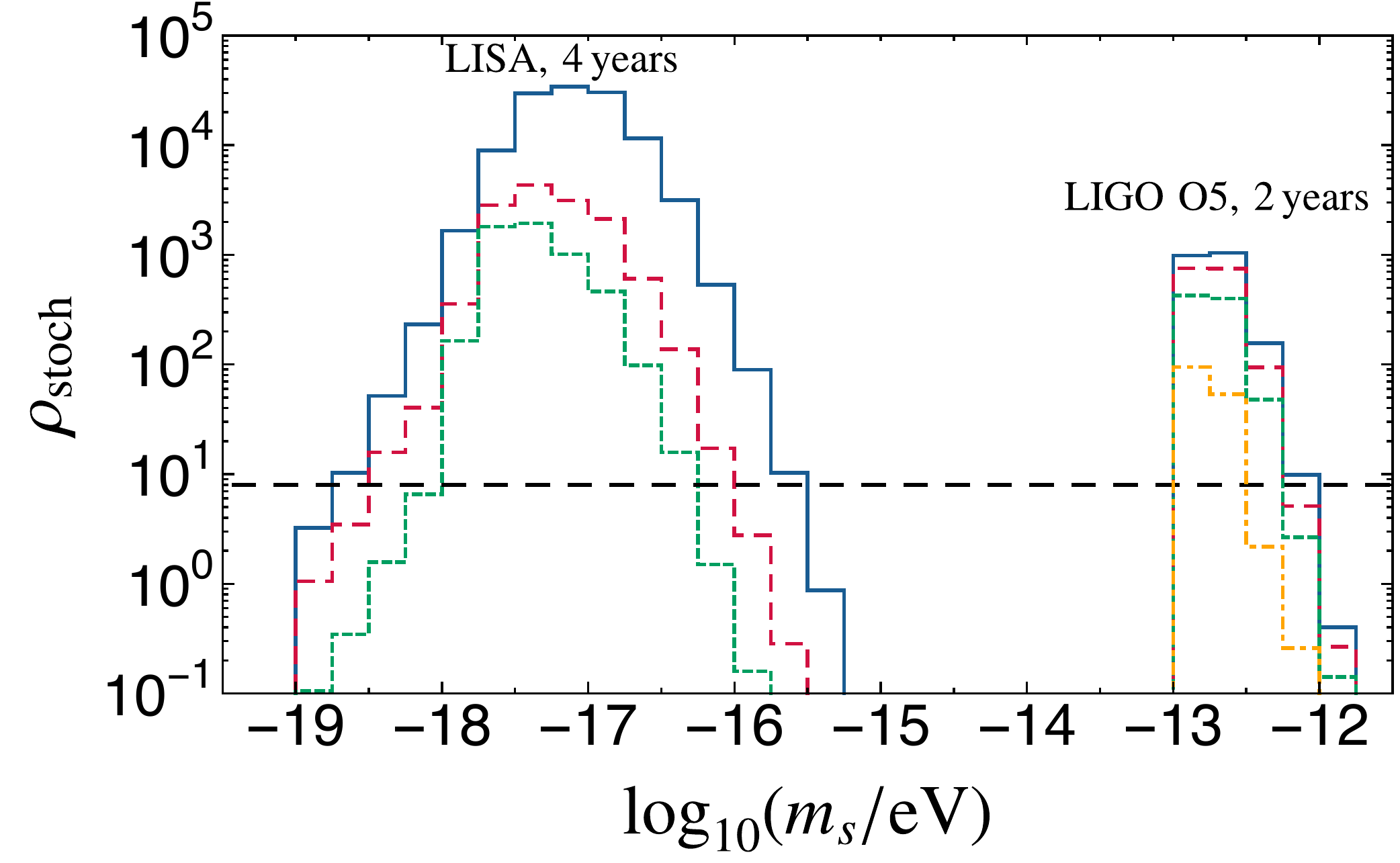,width=0.48\textwidth,angle=0,clip=true}
\caption{Left panel: stochastic background in the LIGO and LISA bands.
  For LISA, the three different signals correspond to the
  ``optimistic'' (top), ``less optimistic'' (middle) and
  ``pessimistic'' (bottom) astrophysical models.  For LIGO, the
  different spectra for each boson mass correspond to a uniform
  spin distribution with (from top to bottom) $\chi_i\in[0.8,1]$,
  $[0.5,1]$, $[0,1]$ and $[0,0.5]$. The black lines are the power-law
  integrated curves of~\cite{Thrane:2013oya}, computed using
  noise PSDs for LISA~\cite{Audley:2017drz}, LIGO's first two
  observing runs (O1 and O2), and LIGO at design sensitivity
  (O5)~\cite{TheLIGOScientific:2016wyq}. By definition,
  $\rho_{\rm stoch}> 1$ ($\rho_{\rm stoch}= 1$) when a power-law spectrum intersects (is tangent to) a power-law integrated curve.  Right panel: $\rho_{\rm stoch}$
  for the backgrounds shown in the left panel. We assumed $T_{\rm
    obs}=2\,{\rm yr}$ for LIGO and $T_{\rm obs}=4\,{\rm yr}$ for LISA.
  \label{fig:background}}
\end{center}
\end{figure*}

The dominant contribution to LIGO resolvable signals comes from
Galactic stellar-mass BHs~\cite{Arvanitaki:2014wva}. We estimate their present-day mass function as
\begin{equation}
\frac{dN_{\rm MW}}{d M}=\int dt\frac{{\rm SFR}(z)}{{\cal M_\star}}\frac{dp}{d{\cal M_{\star}}}\left\vert\frac{dM}{d{\cal M_\star}}\right\vert^{-1}\,, \label{Irina2}
\end{equation}
where the integration is over all cosmic times prior to the present epoch; 
$N_{\rm MW}$ denotes the number of BHs in the Galaxy; ${\rm SFR}(z)$ is the SFR of Milky-Way type galaxies as
a function of redshift~\cite{2013ApJ...770...57B,2013ApJ...762L..31B};
${dp}/{d{\cal M_{\star}}}$ is the
probability of forming a star with mass between ${\cal M_\star}$ and
${\cal M_\star}+d{\cal M_\star}$ (obtained from the Salpeter initial mass function); and $dM/{d{\cal M_\star}}$ is given
by the ``delayed'' model of~\cite{2012ApJ...749...91F}.
This latter quantity is a function of redshift through the metallicity,
whose redshift evolution we model following~\cite{2016MNRAS.456.2140M}.
To obtain a (differential) BH number density ${dn_{\rm MW}}/{d M}$, 
we ``spread'' this mass function over 
the Galaxy, proportionally to the (present) stellar density. For the latter we
assume a simple bulge+disk model, where the bulge is modeled via a
Hernquist profile~\cite{1990ApJ...356..359H} with mass
$\sim 2\times 10^{10} M_\odot$ and scale radius
$\sim 1\,{\rm kpc}$~\cite{Shen:2003sda}, and the disk has an
exponential profile with mass $\sim 6\times 10^{10} M_\odot$ and scale
radius $\sim 2 $ kpc~\cite{1998A&A...330..136P}.

\noindent{{\bf{\em Stochastic background.}}}
The stochastic background produced by BH-boson condensates is given by
an integral over unresolved sources -- those with signal-to-noise
ratio (SNR) $\rho<8$ -- of the formation rate density per comoving
volume $\dot{n}$~\cite{Phinney:2001di}:
\begin{equation}\label{stochastic1}
\Omega_{\rm GW}(f)=\frac{f}{\rho_c}\int_{\rm \rho<8} 
{d} z \frac{{\rm} d t}{{\rm} d z}\\
\dot{n}(M,\chi,z)  \frac{{\rm} d E_s}{{d} f_s}\,,
\end{equation}
where
$\rho_c=3H_0^2/(8\pi)\approx 1.3\times10^{11} M_{\odot}/{\rm Mpc}^3$
is the critical density of the Universe,
${{\rm} d t}/{{\rm} d z}$
is the derivative of the lookback time $t(z)$ with respect to $z$,
${{\rm} d E_s}/{{d} f_s} $ is the energy spectrum in the source frame,
and $f$ is the detector-frame frequency. 
For LIGO we compute $\dot
n$ by integrating Eqs.~(\ref{Irina}) and (\ref{Irina2}).
For LISA we integrate ${d^2 n}/(d \log_{10} M d
\chi)$ --~as given by the aforementioned ``optimistic'', ``less
optimistic'' and ``pessimistic'' models~-- with respect to mass and
spin, and we assume that $\dot n=n/t_0$, where $t_0\approx 13.8\,{\rm
  Gyr}$ is the age of the Universe (i.e., each BH undergoes boson
annihilation only once in its cosmic history). This assumption does
not significantly affect our results, because subsequent annihilation
signals (if they occur at all) are much weaker~\cite{Brito:2017zvb}.

For the spectrum of the GW signal we assume ${{\rm} d E_s}/{{d} f_s}
\approx E_{\rm GW} \delta (f(1+z)-f_s)$, where $E_{\rm
  GW}$ is the total energy radiated in GWs over the signal duration $\Delta t$, and the Dirac delta is
``spread out'' over a frequency window of width 
$\sim \max[1/(\Delta t (1+z)),1/T_{\rm obs}]$ 
 to account for the finite signal duration and observation time.
For LIGO we can safely neglect the effect of
mergers~\cite{Gerosa:2017kvu,Fishbach:2017dwv} and
accretion~\cite{King:1999aq}.
For LISA, we conservatively assume that mergers and accretion cut the
signal short, and thus define the signal duration as
$\Delta t=\left\langle\min\left(\tau_{\rm GW}/(N_m+1),\,t_S,\,t_0\right)\right\rangle$,
where $\tau_{\rm GW}$ is given by Eq.~\eqref{tauGW};
$t_S=4.5\times 10^8 {\rm\, yr\,} {\eta}/[{f_{\rm Edd} (1-\eta)}]$ is
the typical accretion ``Salpeter'' timescale, which depends on
the Eddington ratio $f_{\rm Edd}$  and on the
spin-dependent radiative efficiency $\eta$; $\langle...\rangle$ denotes an
average weighted by the Eddington-ratio probability distribution;
and $N_m$ is the average number of mergers in the interval
$[t(z)-\frac{1}{2}\tau_{\rm GW},t(z)+\frac{1}{2}\tau_{\rm GW}]$~\cite{Brito:2017zvb}.
Moreover, since our calculation assumes that the instability saturates before GW emission takes place, 
our stochastic background calculation only includes BHs for which 
the expected number of mergers during the instability
timescale is $N_m<1$, and for which $\tau_{\rm inst}<\Delta t$ (thus
ensuring that the instability timescale is shorter than the typical accretion and merger timescales).

The ${\rm SNR}$ for the stochastic background is~\cite{Thrane:2013oya}
\be
\rho_{\rm stoch}=\sqrt{T_{\rm obs}\int_{f_{\rm min}}^{f_{\rm max}}df \frac{\Omega^2_{\rm GW}}{\Omega^2_{\rm sens}}}\,,
\ee
where $\Omega^{\rm LIGO}_{\rm sens}=\frac{S_h(f)}{\sqrt{2}\Gamma_{IJ}(f)} \frac{2\pi^2}{3H_0^2}f^3$ and $\Omega^{\rm LISA}_{\rm sens}=S_h(f) \frac{2\pi^2}{3H_0^2}f^3$
for LIGO~\cite{Allen:1997ad} and LISA~\cite{Cornish:2001qi},
respectively. In the LIGO case we assume the same $S_h$ for the
Livingston and Hanford detectors, and $\Gamma_{IJ}$ denotes their
overlap reduction function~\cite{Thrane:2013oya}.

The order of magnitude of the stochastic background shown in
Fig.~\ref{fig:background} (left panel) can be estimated by a simple
back-of-the-envelope calculation.
The average mass fraction of an isolated BH emitted by the boson cloud
is $f_{\rm ax}\sim {\cal O}(1\%)$~\cite{Brito:2017zvb}.
Because the signal is almost monochromatic, the emitted GWs in the
detector frame span about a decade in frequency, i.e.
$\Delta \ln f \sim 1$ for both LISA and LIGO (cf.\
Fig.~\ref{fig:background}).  Thus,
$\Omega_{\rm GW,\,ax} =({1}/{\rho_{\rm c}})({d\rho_{\rm GW}}/{d\ln f})
\sim f_{\rm ax}{\rho_{\rm BH}}/{\rho_{\rm c}}$, where $\rho_{\rm GW}$
and $\rho_{\rm BH}$ are the GW and BH energy density, respectively.
Since the BH mass density is
$\rho_{\rm BH}\sim {\cal O}( 10^4) M_{\odot}/{\rm Mpc}^3$ in the mass
range $10^4-10^7 M_\odot$ relevant for LISA, this yields
$\Omega^{\rm LISA}_{\rm GW,\,ax} \sim 10^{-9}$.  For LIGO, the
background of GWs from BH binaries can be approximated as
$\Omega_{\rm GW,\,bin}\sim f_{\rm GW} f_{\rm m} \rho_{\rm
  BH}/\rho_c$, where $f_{\rm GW}\sim {\cal O}(1\%)$ is the binary's mass
fraction emitted in GWs~\cite{Barausse:2012qz}, and
$f_{\rm m} \sim {\cal O}(1\%)$~\cite{2016MNRAS.461.3877D} is the
fraction of stellar-mass BHs in binaries that merge in less than $t_0$. Therefore
$\Omega_{\rm GW,\,ax}/\Omega_{\rm GW,\,bin}\sim f_{\rm ax}/(f_{\rm GW}
f_{\rm m})\sim 10^2$.  Since the O1 results imply peak background values
$\Omega_{\rm GW,\,bin}\sim 10^{-9} - 10^{-8}$~\cite{TheLIGOScientific:2016wyq,TheLIGOScientific:2016dpb} (or larger if spins
are included), we obtain $\Omega^{\rm LIGO}_{\rm GW,\,ax}\sim 10^{-7}-10^{-6}$. These
estimates are in qualitative agreement with the left panel of
Fig.~\ref{fig:background}.

Remarkably, $\rho_{\rm stoch}$ (right panel of
Fig.~\ref{fig:background}) can be very high. For optimistic
astrophysical models, boson masses in the range
$2\times 10^{-13}\,{\rm eV}\lesssim m_s \lesssim 10^{-12}\,{\rm eV}$
($5\times 10^{-19}\,{\rm eV}\lesssim m_s \lesssim 5\times
10^{-16}\,{\rm eV}$) yield $\rho_{\rm stoch}>8$ with LIGO (LISA).
Our estimates suggest that, for the most pessimistic model and masses around $m_s \approx 3\times 10^{-12}\,{\rm eV}$, the background would have ${\rm SNR}\approx 1.2$ using our simple analytic estimate of the LIGO O1 sensitivity, thus being only marginally allowed by current LIGO O1 upper limits~\cite{TheLIGOScientific:2016dpb}. Our conclusions should be validated by a careful data analysis of the stochastic background in LIGO O1 and O2. In particular, current upper limits on the stochastic background assume that the spectrum can be described by a power law in the LIGO range~\cite{TheLIGOScientific:2016dpb}, which is not the case for the backgrounds computed here.

\noindent{{\bf{\em Resolvable sources.}}}
We estimate the number of resolvable events as~\cite{Brito:2017zvb}
\begin{equation}\label{RATES}
{N}=  \int_{{\rho} >8}\frac{d^2 \dot{n}}{d M  d \chi} \left(\frac{T_{\rm obs}}{1+z}+\Delta t\right) \frac{d V_c}{dz}dz d M d \chi \,,
\end{equation}
where $dV_c=4 \pi D_c^2 {d D_c}$, $D_c$ is the comoving distance, and
$\dot n=n/t_0$ for LISA.
The dependence on $T_{\rm obs}/(1+z)+\Delta t$ comes about because the
probability that an observation of duration $T_{\rm obs}$ and a signal
of duration $\Delta t (1+z)$ (in the detector frame) overlap is
proportional to the sum of the two durations. In the limit
$\Delta t (1+z)\ll T_{\rm obs}$ we have $N\propto T_{\rm obs}$, as
usual for short-lived sources~\cite{Hartwig:2016nde}. For
$\Delta t (1+z)\gg T_{\rm obs}$, $N$ becomes proportional to the duty
cycle $\Delta t/t_f$, $t_f\equiv n/\dot{n}$ being the formation
timescale of the boson condensates. This duty cycle, akin e.g. to the
duty cycle of active galactic nuclei, accounts for the fact that only
a fraction of the sources are radiating during the observation time.
\begin{figure}[htb]
\begin{center}
\epsfig{file=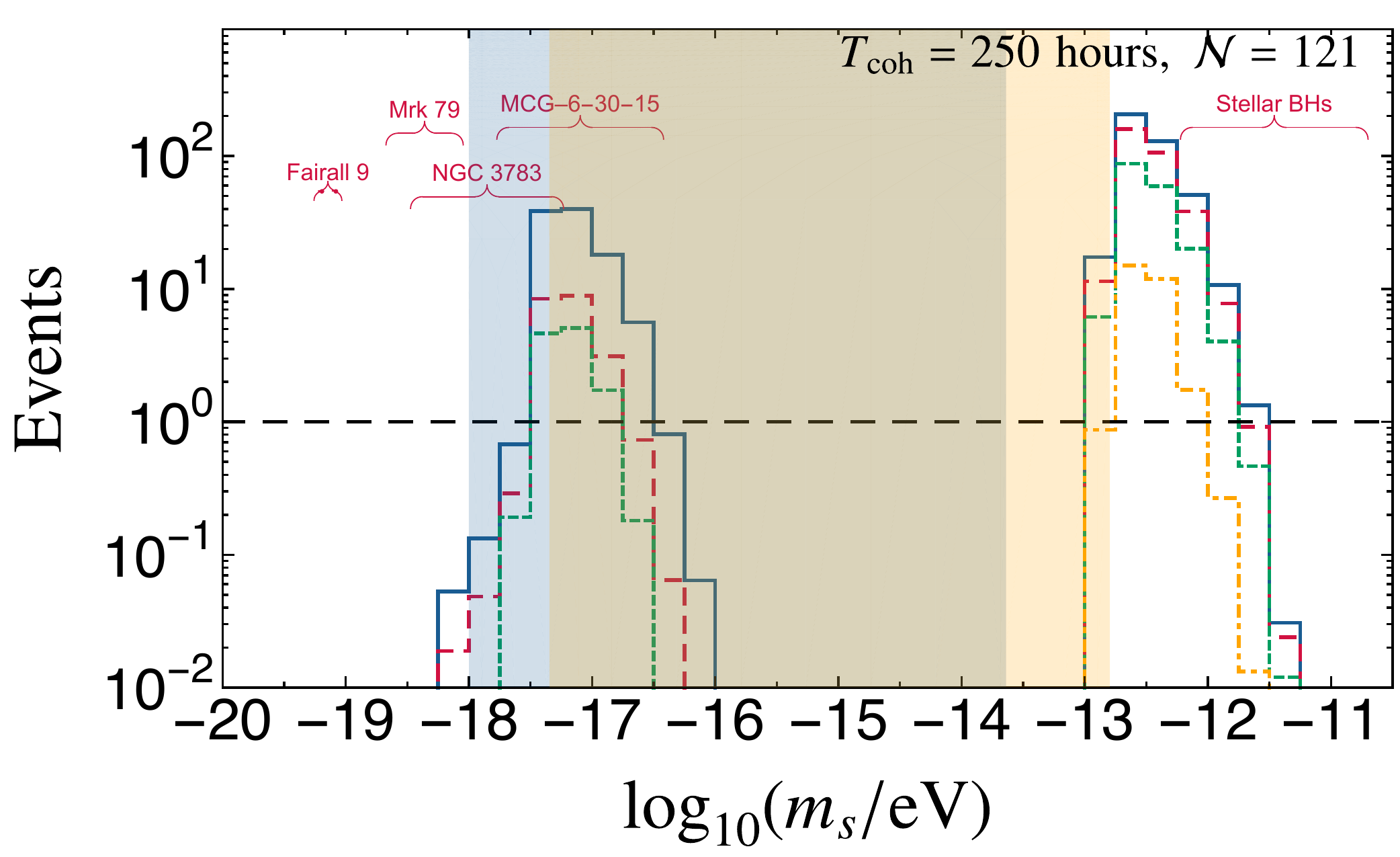,width=0.48\textwidth,angle=0,clip=true}
\caption{Resolvable events for the same astrophysical models used in
  Fig.~\ref{fig:background}. Shaded areas correspond to exclusion
  regions from 4-year LISA massive BH spin measurements, using either
  the ``popIII'' (brown) or ``Q3-nod'' (light blue) models
  of~\cite{Klein:2015hvg}.  For reference we also show with brackets the constraints
  that can be placed by spin measurements of massive/stellar-mass BHs~\cite{Brenneman:2011wz,Arvanitaki:2014wva}.
  \label{fig:resolved}}
\end{center}
\end{figure}

Figure~\ref{fig:resolved} shows resolvable event rates assuming
(conservatively) semicoherent searches for different astrophysical
models.
The SNR was computed by including the confusion noise from the
stochastic background of unresolvable boson-condensate sources:
neglecting this contribution would overestimate LISA rates by more
than one order of magnitude.
Our models typically predict $\sim$ 40 (200) events in $121\times 250$
hours of total observation time for the optimistic models and boson
masses in the optimal range around $m_s\sim 10^{-17}\,{\rm eV}$
($3\times 10^{-13}\,{\rm eV}$) for LISA (LIGO).
Rates in the less optimistic and pessimistic models decrease by
factors of order unity.  However, it is remarkable that a boson with
$m_s\sim 10^{-17}\,{\rm eV}$ ($3\times 10^{-13}\,{\rm eV}$) would
produce around $5$ ($15$) direct LISA (LIGO) detections even for
pessimistic astrophysical models.

So far we focused on the {\em direct} detection of GWs from bosonic
condensates. In~\cite{Brito:2017zvb} we use Bayesian model selection to
show that LISA could infer the existence of light bosons {\it
  indirectly}: LISA measurements of massive BH spins could provide
evidence for holes in the BH mass-spin ``Regge plane'' (i.e., for the
absence of BHs spinning above the superradiant instability
window)~\cite{Arvanitaki:2010sy}. As indicated by the shaded areas in
Fig.~\ref{fig:resolved}, a 4-year LISA mission could rule out boson
masses in a range that depends on the assumed BH model
($[4.5 \times 10^{-18}, 1.6 \times 10^{-13}]$ for the ``light-seed''
popIII model, $[10^{-18}, 2.3 \times 10^{-14}]$ for the
``heavy-seed, no-delay'' Q3-nod model of~\cite{Klein:2015hvg}). If
fields with $m_s\in [10^{-17}, 10^{-13}]$~eV exist in nature, LISA
observations of BH mergers can measure $m_s$ with ten percent
accuracy~\cite{Brito:2017zvb}.

\noindent{{\bf{\em Conclusions.}}}
%
Together, Earth- and space-based detectors will allow for multiband GW
searches of ultralight bosons in the range
$[10^{-19}-10^{-10}]\,{\rm eV}$. We plan to improve estimates of the
stochastic background for LIGO by using population synthesis
models~\cite{Berti:2016lat,Belczynski:2016jno,Belczynski:2016ieo}.
The potential of detectors like DECIGO or the Einstein Telescope to
detect or rule out bosons of mass $m_s\sim 10^{-14} \,{\rm eV}$ should also be
investigated by using intermediate-mass BH formation
models~\cite{Kawamura:2006up,Gair:2010dx}. Our analysis must be
extended to spin-1~\cite{Pani:2012vp,Pani:2012bp} and
spin-2~\cite{Brito:2013wya} fields, for which the instability time
scales are shorter and GW amplitudes are larger. Our results also
suggest that recent estimates of resolvable GWs from spin-1
instabilities~\cite{Baryakhtar:2017ngi} should be revised taking into
account the stochastic background.

\noindent{\bf{\em Acknowledgments.}}
%
We would like to thank Lijing Shao for useful comments.
S.~Ghosh and E.~Berti are supported by NSF Grants No.~PHY-1607130 and
AST-1716715. E.~Berti is supported by FCT contract
IF/00797/2014/CP1214/CT0012 under the IF2014 Programme.
V.~Cardoso acknowledges financial support provided under the European Union's H2020 ERC Consolidator Grant ``Matter and strong-field gravity: New frontiers in Einstein's theory'' grant agreement no. MaGRaTh--646597.
Research at Perimeter Institute is supported by the Government of Canada through Industry Canada and by the Province of Ontario through the Ministry of Economic Development $\&$
Innovation.
The work of I.~Dvorkin has been done within the Labex ILP (reference ANR-10-LABX-63) part of the Idex SUPER, and received financial state aid managed by the Agence Nationale de la Recherche, as part of the programme Investissements d'avenir under the reference ANR-11-IDEX-0004-02.
This project has received funding from the European Union's Horizon 2020 research and innovation programme under the Marie Sklodowska-Curie grant agreement No 690904 and from FCT-Portugal through the project IF/00293/2013.
The authors would like to acknowledge networking support by the COST Action CA16104.
This work has made use of the Horizon Cluster, hosted by the Institut
d'Astrophysique de Paris. We thank Stephane Rouberol for running
smoothly this cluster for us.
The authors thankfully acknowledge the computer resources, technical expertise and assistance provided by S\'ergio Almeida at CENTRA/IST. Computations were performed at the cluster ``Baltasar-Sete-S\'ois'', and supported by the MaGRaTh--646597 ERC Consolidator Grant.
This work was supported by the French Centre National d'{\'E}tudes Spatiales (CNES).

%


\bibliographystyle{apsrev4}
\bibliography{Ref}

\end{document}